\documentclass[%
reprint,
amsmath,amssymb,
aps,
longbibliography,
]{revtex4-2}

\usepackage{pgffor}
\usepackage{float}
\usepackage{dcolumn}
\usepackage{bm}
\usepackage{amsmath}
\usepackage{mathtools}
\usepackage[T1]{fontenc}
\usepackage{xspace}
\usepackage{comment}
\usepackage{bbold}
\usepackage{braket}
\usepackage{ascmac}
\usepackage{graphicx}
\usepackage{hyperref}
\usepackage{ulem}

\makeatletter
\let\MYcaption\@makecaption
\makeatother
\usepackage{subcaption}
\makeatletter
\let\@makecaption\MYcaption
\makeatother

\ifx\pdfoutput\undefined
\usepackage[dvipdfmx]{graphicx}
\usepackage[dvipdfmx]{hyperref}
\usepackage[dvipdfmx]{color}
\usepackage[dvipdfmx]{xcolor}
\else
\usepackage{graphicx}
\usepackage{hyperref}
\usepackage{color}
\usepackage{xcolor}
\fi

\begin{document}

\title{Spin-liquid and spin-glass behavior in quantum spin models\\with all-to-all $p$-spin interactions}
\author{Shusei Wadashima}
\author{Yukitoshi Motome}
\affiliation{
 Department of Applied Physics, The University of Tokyo, Hongo, Tokyo 113-8656, Japan
}
\date{\today}
\begin{abstract}
Spin-liquid and spin-glass states represent two distinct phases of disordered quantum spin systems. These states are, in principle, distinguished by quantum-entangled fluctuations and spin freezing, but identifying each state and characterizing the transition between them remains challenging. Here, we systematically explore the relationship between the spin-liquid and spin-glass states using a model with all-to-all random interactions among $p$ spins, which interpolates between the Ising-like one-component, $XY$-like two-component, and isotropic three-component cases. By analyzing the system-size $N$ dependence of the Edwards-Anderson order parameter and the density of states, we identify the transition from the spin liquid to the spin glass for various values of $p$. We show that the phase diagrams for different $p$ can be unified through a scaling with $N/p^2$, revealing that increasing anisotropy in the interactions systematically suppresses the spin-liquid phase and extends the spin-glass regime. Furthermore, we examine the competition between multiple-spin interactions and anisotropy under an external magnetic field in the isotropic case, and find that the spin-liquid phase transitions into the spin-glass phase before entering a quantum paramagnetic phase. Our findings provide insights into quantum disordered phases and the transitions between them.
\end{abstract}
\maketitle

\section{Introduction\label{sec:introduction}}
When the temperature is lowered, most magnets exhibit long-range magnetic orders, such as ferromagnetic and antiferromagnetic orders, accompanied by spontaneous symmetry breaking. However, when the interactions between magnetic moments are frustrated and compete with each other, long-range ordering can be suppressed to lower temperatures, and in extreme cases, even down to absolute-zero temperature. Notable examples of these exceptions are quantum spin liquid and spin glass. The quantum spin liquid is a quantum disordered state where the magnetic moments are strongly fluctuating and quantum entangled with each other~\cite{Balents2010, Savary2017}. In contrast, the spin glass is also a disordered state, but one in which the magnetic moments exhibit freezing in random orientations with less entanglement among them~\cite{Binder1986, Mydosh1993}. Thus, these two states are fundamentally distinct phases of magnets in terms of quantum fluctuations, entanglement, and freezing. Nevertheless, identifying each phase and characterizing the phase transition between them remains a nontrivial task, as both share a disordered nature and do not have conventional order parameters.

The concept of quantum spin liquid was initiated by Anderson's pioneering work, where he proposed the resonating valence-bond state for the ground state of triangular-lattice antiferromagnets with spin $S=1/2$ moments~\cite{Anderson1973}. This state is defined by a superposition of spin-singlet pairs covering the entire lattice, without breaking any symmetry of the system. Although subsequent studies revealed that the true ground state exhibits a long-range order with a noncollinear spin configuration, Anderson's proposal sparked significant progress in the study of quantum spin liquids, including the discovery of fractional excitations~\cite{Read1989, Wen1991} and topological order~\cite{Wen1989}. The resonating valence-bond state has also been explored as a possible mechanism underlying high-temperature superconductivity~\cite{Anderson1987}. Despite these advances, the research field has faced a major challenge: the lack of realistic model Hamiltonians that realize quantum spin liquids in their ground state. This challenge was addressed by the proposal of the Kitaev model, whose ground state is exactly solvable and shown to be a quantum spin liquid~\cite{Kitaev2006}. The ground state of the Kitaev model hosts fractional excitations in the form of Majorana fermions and anyons, and exhibits topological order. This discovery and subsequent developments have not only deepened our understanding of quantum spin liquids but also paved the way for their realization in materials and the exploration of topological quantum computing utilizing these fractional excitations~\cite{Takagi2019,Motome2020,Trebst2022}. 

Meanwhile, the study of spin glass gained momentum following its experimental discovery in dilute magnetic alloys by Cannella and Mydosh in 1972~\cite{Cannella1972}. Theoretical progress accelerated with the introduction of the Edwards-Anderson (EA) model, which features random two-spin interactions between the Ising spins~\cite{Edwards1975}. It was proposed that the spin-glass state can be characterized by the overlap measure between spin configurations observed at different times, known as the EA order parameter~\cite{Edwards1975}. Further development came with the analytical solution of the Sherrington-Kirkpatrick (SK) model, which also involves random two-spin interactions, but between all pairs of spins (all-to-all coupling), with coupling strengths drawn from the Gaussian distribution~\cite{Sherrington1975}. The replica method provided the exact solution for this model, highlighting a phase transition from a high-temperature paramagnetic phase to a low-temperature spin-glass phase. In addition, the concept of replica symmetry breaking was introduced in the context of this model by Parisi~\cite{Parisi1979}, which deepened the understanding of spin-glass states by clarifying the existence of a complex energy landscape with numerous energy minima. The SK model was later generalized to include higher-order interactions among multiple spins~\cite{Derrida1980} and quantum fluctuation effects~\cite{Bray1980}, both of which preserve spin-glass nature.

In recent years, randomness has come to play a central role not only in freezing phenomena like spin glass but also in quantum disordered states without freezing like spin liquid. The Sachdev-Ye-Kitaev (SYK) model~\cite{Kitaev2015, Polchinski2016, Maldacena2016D, Kitaev2018, Trunin2021} provides a notable theoretical framework for exploring such unconventional states. This model describes fermions with all-to-all random couplings, leading to a quantum disordered phase characterized by emergent properties, such as maximal chaos~\cite{Maldacena2016H}, conformal symmetry in the low-energy limit~\cite{Kitaev2015, Maldacena2016D}, and the absence of quasiparticle excitations~\cite{Sachdev2015, Maldacena2016D, Davison2017}. Although the SYK model shares all-to-all random interactions with the SK model, it does not exhibit freezing behavior like spin glass~\cite{Gur-Ari2018, Wang2019, Arefeva2019}. Due to these distinctive characteristics, the SYK model has attracted considerable interest across various fields, including condensed matter physics~\cite{Chowdhury2022, Hartnoll2018}, quantum chaos~\cite{Cotler2017, Garcia-Garcia2017, Garcia-Garcia2018}, and quantum gravity~\cite{Jensen2016}. The model has been generalized to quantum spin systems, highlighting spin-liquid behavior in the intermediate frequency and temperature ranges~\cite{Shackleton2021, Christos2022}. Thus, the SYK model and its extensions have brought new perspectives on spin-liquid and spin-glass states. Nevertheless, the relationship and distinction between these two states remain elusive despite their importance in exploring the fundamental properties of quantum disordered systems.

In this study, we investigate a transition between spin-liquid and spin-glass states induced by randomness in quantum spin systems. Spin models, which possess complexity arising from random quantum interactions, may exhibit either freezing or nonfreezing behavior, depending on the nature of the interactions, such as their structural complexity and degree of quantumness. To address this issue, we study a spin $S=1/2$ model with all-to-all random interactions involving multiple $p$ spins, which enables a systematic exploration by tuning the interaction complexity and anisotropy in spin space. To characterize the nature of the emergent phases, we systematically analyze the system-size $N$ dependence of the EA order parameter and the density of states, while varying $p$ and the interaction anisotropy in spin space. Our results reveal a crossover from spin-liquid to spin-glass behavior with increasing $N$ for each $p$, with the crossover point being highly sensitive to the interaction anisotropy. We elucidate the phase diagram through scaling with $N/p^2$, suggesting that the spin-liquid phase is rapidly suppressed with increasing the interaction anisotropy, and ultimately vanishes in the Ising limit. Furthermore, by examining the effect of an external magnetic field in the isotropic case, we find that the spin-liquid phase observed in the absence of a magnetic field transitions into the spin-glass phase before entering a quantum paramagnetic phase.

The structure of this paper is as follows. Section~\ref{sec:model} introduces the model employed in this study and demonstrates how it encompasses several limiting cases previously investigated in the literature. Section~\ref{sec:method} outlines the method and defines the physical quantities used for the analysis. Section~\ref{sec:results} presents the results for the behavior of spin-liquid and spin-glass states under various interaction regimes: the isotropic limit (Sec.~\ref{sec:isotropic}), the $XY$ anisotropic limit (Sec.~\ref{sec:XY}), and the Ising limit (Sec.~\ref{sec:Ising}). The findings are summarized in a phase diagram in Sec.~\ref{sec:diagram}. Effects of an external magnetic field are explored in Sec.~\ref{sec:field}. Finally, Sec.~\ref{sec:conclusion} summarizes the key findings of this study and discusses potential directions for future research. The Appendix provides the derivation of the expression for the density of states in the thermodynamic limit, which is used to discuss the phase diagram in Sec.~\ref{sec:diagram}.

\section{Model\label{sec:model}}
We study a spin model with random all-to-all interactions in a uniform external magnetic field, whose Hamiltonian is given by
\begin{align}
H = \sum_{\substack{1 \le i_1 <  \dots < i_p \le N \\  \alpha_1, \dots, \alpha_p \in \{x, y, z\}}} \frac{J_{i_1 \alpha_1 \dots i_p \alpha_p}}{\sqrt{\mathcal{N}_{N,p}}} \tilde{\sigma}_{i_1}^{\alpha_1} \dots \tilde{\sigma}_{i_p}^{\alpha_p}
+ h \sum_{i} \tilde{\sigma}_i^z,
\label{eq:H_general}
\end{align}
with
\begin{align}
\mathcal{N}_{N,p} = \binom{N}{p} \left( 1 + \varepsilon_x^2 + \varepsilon_y^2 \right)^p.
\end{align}
The first term denotes the random all-to-all interactions, each of which involves $p$ spins among total $N$ spins; the coupling constants \( J_{i_1 \alpha_1 \dots i_p \alpha_p} \) are assumed to be independent and identically distributed Gaussian random variables with zero mean and unit variance. The interaction is anisotropic in spin space, which is controlled by the parameters $\varepsilon_x$ and $\varepsilon_y$ as
\begin{align}
\tilde{\sigma}^x = \varepsilon_x \sigma^x, \quad \tilde{\sigma}^y = \varepsilon_y \sigma^y, \quad \tilde{\sigma}^z = \sigma^z,
\label{eq:sigma_tilde}
\end{align}
where \(0 \leq \varepsilon_x, \varepsilon_y \leq 1\) and $(\sigma^x,\sigma^y,\sigma^z)$ represent the Pauli matrices. The second term in Eq.~\eqref{eq:H_general} denotes the Zeeman coupling with the external magnetic field $h$ applied in the $z$ direction.

\begin{figure}[tb]
\centering 
\includegraphics[width=\columnwidth]{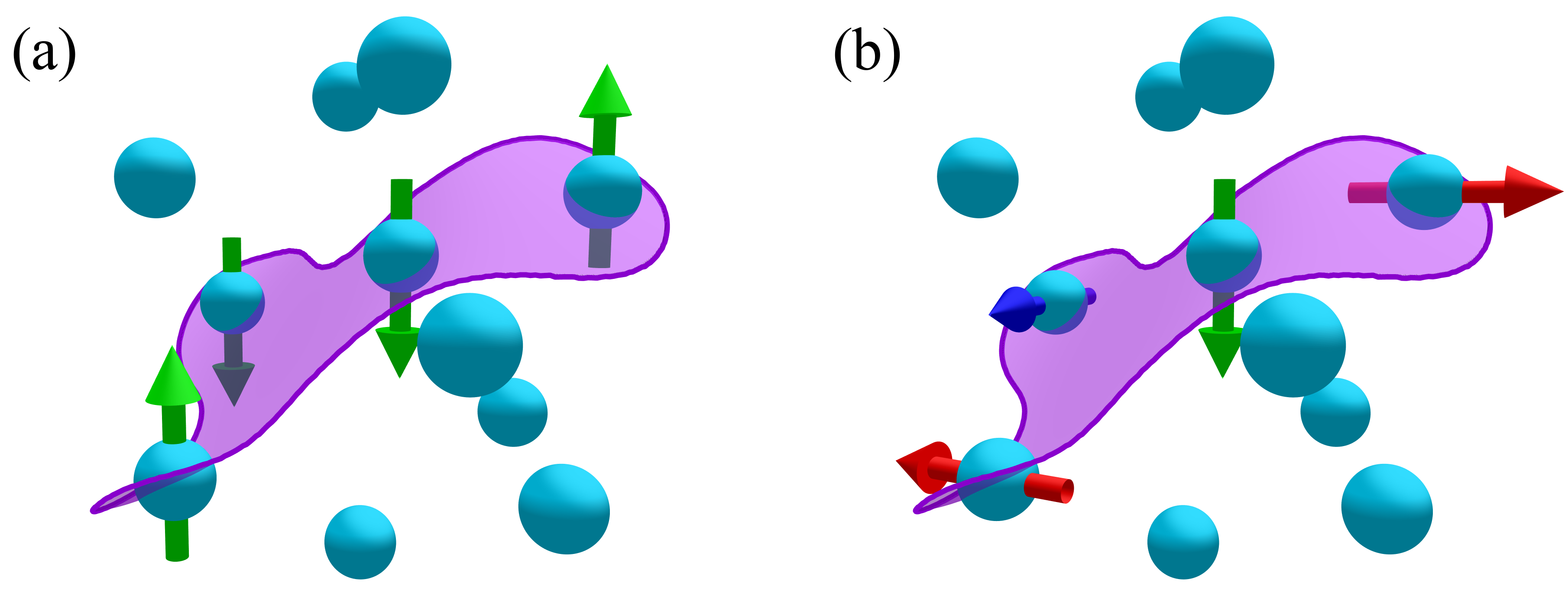} 
\caption{Schematic illustrations of $p$-spin interactions in the model in Eq.~\eqref{eq:H_general} for two cases: (a) \((\varepsilon_x, \varepsilon_y) = (0, 0)\), corresponding to an Ising-like one-component model and (b) \((\varepsilon_x, \varepsilon_y) = (1, 1)\), corresponding to an isotropic three-component model. Blue spheres represent $N$ sites, and colored arrows indicate spins involved in a $p$-spin interaction highlighted by the purple region in each illustration ($p=4$ in this case). Spins are defined on all blue spheres, but for clarity, they are explicitly shown only in the purple-highlighted region.}
\label{fig:schema}
\end{figure}

Figure~\ref{fig:schema} illustrates schematic pictures of the interaction in Eq.~\eqref{eq:H_general} for two extreme cases: (a) \((\varepsilon_x, \varepsilon_y) = (0, 0)\) and (b) \((\varepsilon_x, \varepsilon_y) = (1, 1)\). The former describes the model only with the $z$-spin component, i.e., an Ising-type classical model, referred to as the classical $p$-spin model. The latter corresponds to a quantum model with isotropic interactions, referred to as the isotropic $p$-spin model. The classical $p$-spin model was introduced as a generalization of the SK model with higher-order interactions~\cite{Derrida1980}, and is known to exhibit spin-glass behavior~\cite{Derrida1980, Gardner1985, Kirkpatrick1987}. Meanwhile, the isotropic $p$-spin model was introduced in Ref.~\cite{Erdos2014} and manifests a classical-quantum transition in the density of states at the scale of \(p \sim N^{1/2} \)~\cite{Erdos2014}. The isotropic $p$-spin model has also been studied as an extension of the SYK model to spin systems~\cite{Berkooz2018, Berkooz2019}. It was shown that the low-temperature regime cannot be described by the annealed free energy, suggesting that the model potentially exhibits spin-glass behavior at low temperatures~\cite{Baldwin2020}. Numerical studies have also shown that the model exhibits spin-glass behavior for $p=2$, $3$, and likely $4$, and a SYK-like spin-liquid behavior for \( p \geq 5 \)~\cite{Swingle2024}. However, the system-size dependence of this behavior has not been fully clarified, nor has its relation to the density of states been discussed, presumably because of the limitation of numerical calculations. In addition, our model includes a two-component quantum spin model with \( (\varepsilon_x, \varepsilon_y) = (1, 0) \) or \( (0, 1) \), later referred to as the $XY$ $p$-spin model. This model with \( p = 4 \) was studied in comparison with the SYK model, and quantitative similarities in the density of states, the spectral form factor, and the two-point functions, among others, were observed~\cite{Hanada2024}.

Our model in Eq.~\eqref{eq:H_general} includes all these cases previously studied. In the following, we study the effect of spin anisotropy by changing the parameters \( (\varepsilon_x, \varepsilon_y) \). Based on the systematic study over \(N\) and \(p\) by extensive numerical simulations, we elucidate the phase diagram including spin-liquid and spin-glass states. Interestingly, even for the isotropic $p$-spin model that was previously challenging to access directly, we uncover a transition between the two states. Furthermore, under an external magnetic field, the system tends to turn into a paramagnetic state with forced ferromagnetic moments. We therefore discuss the transitions among spin-glass, spin-liquid, and paramagnetic states within the model in Eq.~\eqref{eq:H_general} in a unified manner.

\section{Method\label{sec:method}}
We investigate the eigenstates of the model in Eq.~\eqref{eq:H_general} using the exact diagonalization of the Hamiltonian defined on clusters with $N$ sites, based on LAPACK with complex double precision. We consider the number of spins involved in the interactions ranging from $p = 3$ to $6$, with system size ranging from $N = p+2$ to $13$ for $p=3$, and from $N = p+1$ to 13 for $p=4$, $5$, and $6$. The results are averaged over random realizations of the coupling constants \( J_{i_1 \alpha_1 \dots i_p \alpha_p} \); the number of random samples will be shown for each case.

To identify the nature of the ground state, we compute the EA order parameter and the density of states. The EA order parameter is used to assess spin-glass behavior, taking a nonzero value in glass phases and vanishing in nonglass phases in the large-\( N \) limit. While it is usually calculated using only the ground state, in this study, we adopt a form that uses the two lowest-energy states as
\begin{align}
q_{\mathrm{EA}} = \frac{1}{2N} \sum_{i,\alpha} \sum_{a,b=1,2} \left| \langle \psi_a | \sigma_i^\alpha | \psi_b \rangle \right|^2,
\label{eq:qEA}
\end{align}
where $a$ and $b$ label the two states in the same sample. This is because the model in Eq.~\eqref{eq:H_general} has a unique property depending on \( p \). For even \( p \), the system has an antiunitary time-reversal symmetry, leading to the ground-state degeneracy that depends on $N$: when $N$ is even, the ground state is nondegenerate, whereas for odd $N$, the ground state has twofold degeneracy in accordance with Kramers' theorem. Meanwhile, for odd \( p \), the system breaks time-reversal symmetry, and the ground state shows no degeneracy regardless of $N$~\cite{Swingle2024}. The definition in Eq.~\eqref{eq:qEA} allows us to study the phase diagram in a unified manner, independent of the parity of \( N \) and \( p \). 

The density of states is defined as 
\begin{align}
\rho(E) = \frac{1}{2^N} \sum_{i} \delta(E - E_i),
\label{eq:dos}
\end{align}
where $E_i$ is the $i$th energy eigenvalue. We numerically calculate this quantity for finite-size systems and present the results as a histogram with $501$ bins, whose centers are evenly spaced from $-3$ to $3$. The density of states in the thermodynamic limit of \(N\to \infty\) with \( p^2/N \) fixed for the isotropic $p$-spin model is found in Ref.~\cite{Erdos2014}, and its generalization to the model in Eq.~\eqref{eq:H_general} with \( h = 0 \) is discussed in the Appendix.

\section{Results\label{sec:results}}
\begin{figure}[tb]
\centering
\includegraphics[width=0.6\columnwidth]{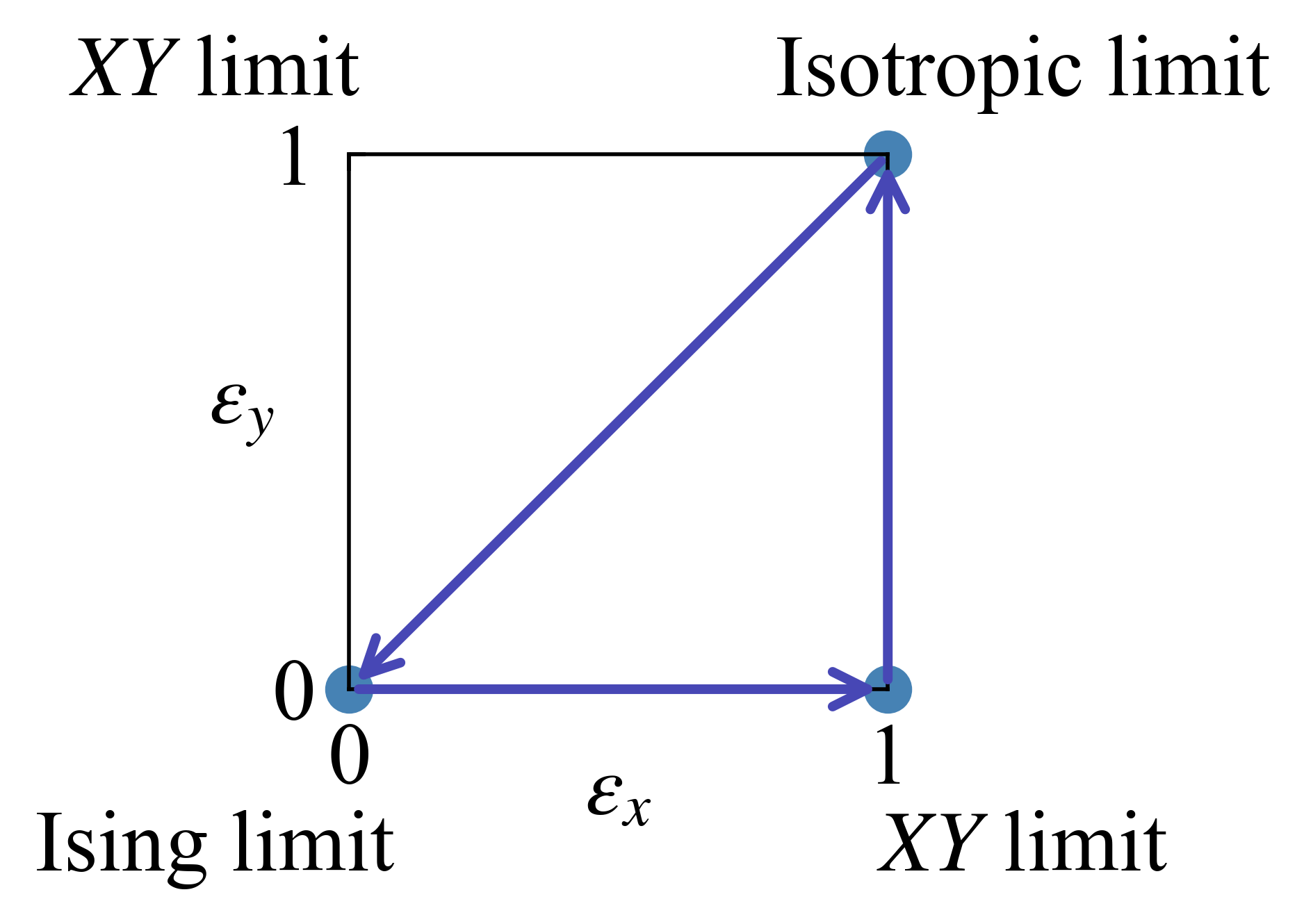}
\caption{Setup of $(\varepsilon_x, \varepsilon_y)$. The cases $(\varepsilon_x, \varepsilon_y) = (1, 1)$, $(1, 0)$, and $(0, 0)$ are discussed in Secs.~\ref{sec:isotropic}–\ref{sec:Ising}, respectively. The connections between these points, represented by the three arrows, are analyzed in Sec.~\ref{sec:diagram}.}
\label{fig:epsilon}
\end{figure}

The analysis begins with three specific cases: \((\varepsilon_x, \varepsilon_y) = (1, 1)\) (Sec.~\ref{sec:isotropic}), \((\varepsilon_x, \varepsilon_y) = (1, 0)\) (Sec.~\ref{sec:XY}), and \((\varepsilon_x, \varepsilon_y) = (0, 0)\) (Sec.~\ref{sec:Ising}), with \(h = 0\); see Fig.~\ref{fig:epsilon}. The first and third cases correspond to the isotropic $p$-spin model in Fig.~\ref{fig:schema}(b) and the classical $p$-spin model in Fig.~\ref{fig:schema}(a), respectively. In the first case, due to the isotropic random quantum interactions for multiple spins, we discover both spin-liquid and spin-glass behavior and a transition between them by changing $N$ and $p$. On the contrary, for the third case, the classical Ising-type interactions produce only spin glass behavior, being consistent with the previous studies~\cite{Derrida1980, Gardner1985, Kirkpatrick1987}. In the second case with anisotropic $XY$-like interactions (although it is technically $XZ$ anisotropic, this is equivalent to $XY$ due to symmetry), both spin-liquid and spin-glass behavior appear similar to the first isotropic case, but the conditions for observing spin-liquid behavior become more restricted than the first case. In Sec.~\ref{sec:diagram}, we elucidate the phase diagram including these three cases while changing \((\varepsilon_x, \varepsilon_y)\) along the blue arrows in Fig.~\ref{fig:epsilon}, and clarify the conditions under which spin-liquid or spin-glass behavior appears. In Sec.~\ref{sec:field}, the analysis is extended by applying an external magnetic field, where we discuss how the system changes from the spin-liquid and spin-glass to a paramagnetic state with increasing the magnetic field.

\subsection{Spin-liquid and spin-glass behavior in the isotropic limit\label{sec:isotropic}}
\begin{figure}[tb]
\centering
\includegraphics[width=\columnwidth]{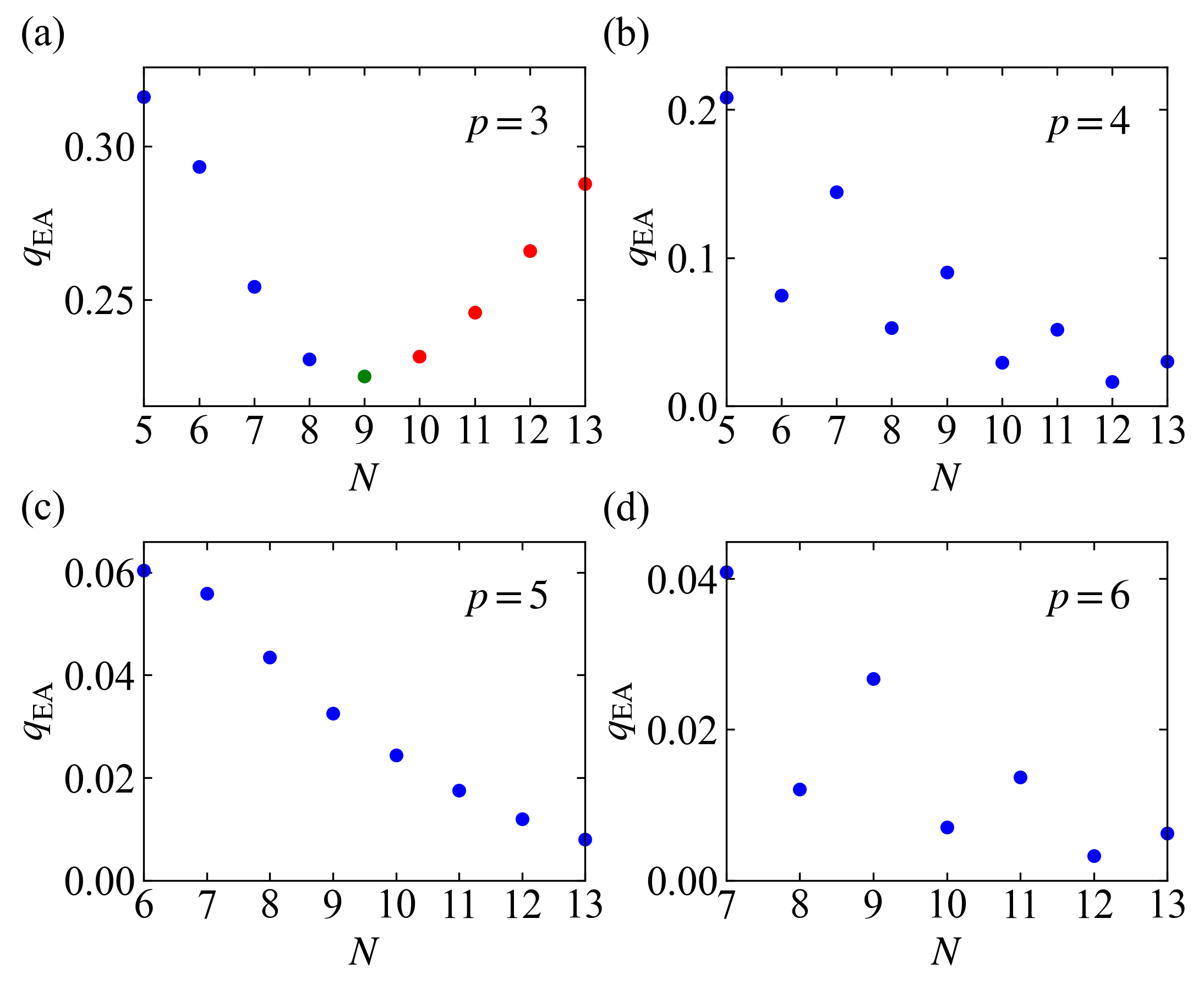}
\vspace{-20pt}
\caption{EA order parameter \( q_{\mathrm{EA}} \) 
for the isotropic $p$-spin model with \((\varepsilon_x, \varepsilon_y) = (1, 1)\) for (a) \( p = 3 \), (b) \( p = 4 \), (c) \( p = 5 \), and (d) \( p = 6 \). The blue, green, and red dots indicate decreasing trends, minima, and increasing trends with respect to $N$, respectively. For even \( p \), these trends are determined separately for even and odd \( N \). The data are obtained by averaging over 300 random samples.}
\label{fig:qEA_isotropic}
\end{figure}

First, we present the results for the isotropic $p$-spin model at zero magnetic field, namely, the Hamiltonian in Eq.~\eqref{eq:H_general} with \((\varepsilon_x, \varepsilon_y) = (1, 1)\) and \( h = 0\). Figure~\ref{fig:qEA_isotropic} shows the $N$ dependence of the EA order parameter \( q_{\mathrm{EA}} \) for $p=3$, $4$, $5$, and $6$. We find that for odd $p$, \( q_{\mathrm{EA}} \) changes smoothly with $N$, while for even $p$, it exhibits an oscillation between even and odd $N$. This is presumably because of the ground-state degeneracy mentioned in Sec.~\ref{sec:method}. Similar behavior was found in Ref.~\cite{Swingle2024}. Importantly, the increase in \( q_{\mathrm{EA}} \) with $N$ can be regarded as a hallmark of instability toward a spin-glass state, which is seen in the case of \( p = 3 \) for $N\geq 9$ in Fig.~\ref{fig:qEA_isotropic}(a). Meanwhile, the decrease in \( q_{\mathrm{EA}} \) with $N$ signifies the absence of spin-glass order, which is observed for \( p = 3 \) in the smaller \( N \) region as well as for the other $p$ in the entire $N$ region. The latter can be interpreted as a tendency toward a spin-liquid state caused by the complexity of the random multiple-spin interactions. Thus, we employ the system size $N$ minimizing \( q_{\mathrm{EA}} \), as a marker to distinguish between the spin-liquid and spin-glass states, for instance, $N=9$ at $p=3$. By this definition, in this isotropic case with \((\varepsilon_x, \varepsilon_y) = (1, 1)\), the spin-glass behavior is found only for \( p = 3 \) and \( N > 9 \), and all the other cases are classified into the spin-liquid state.

\begin{figure}[tb]
\centering
\includegraphics[width=\columnwidth]{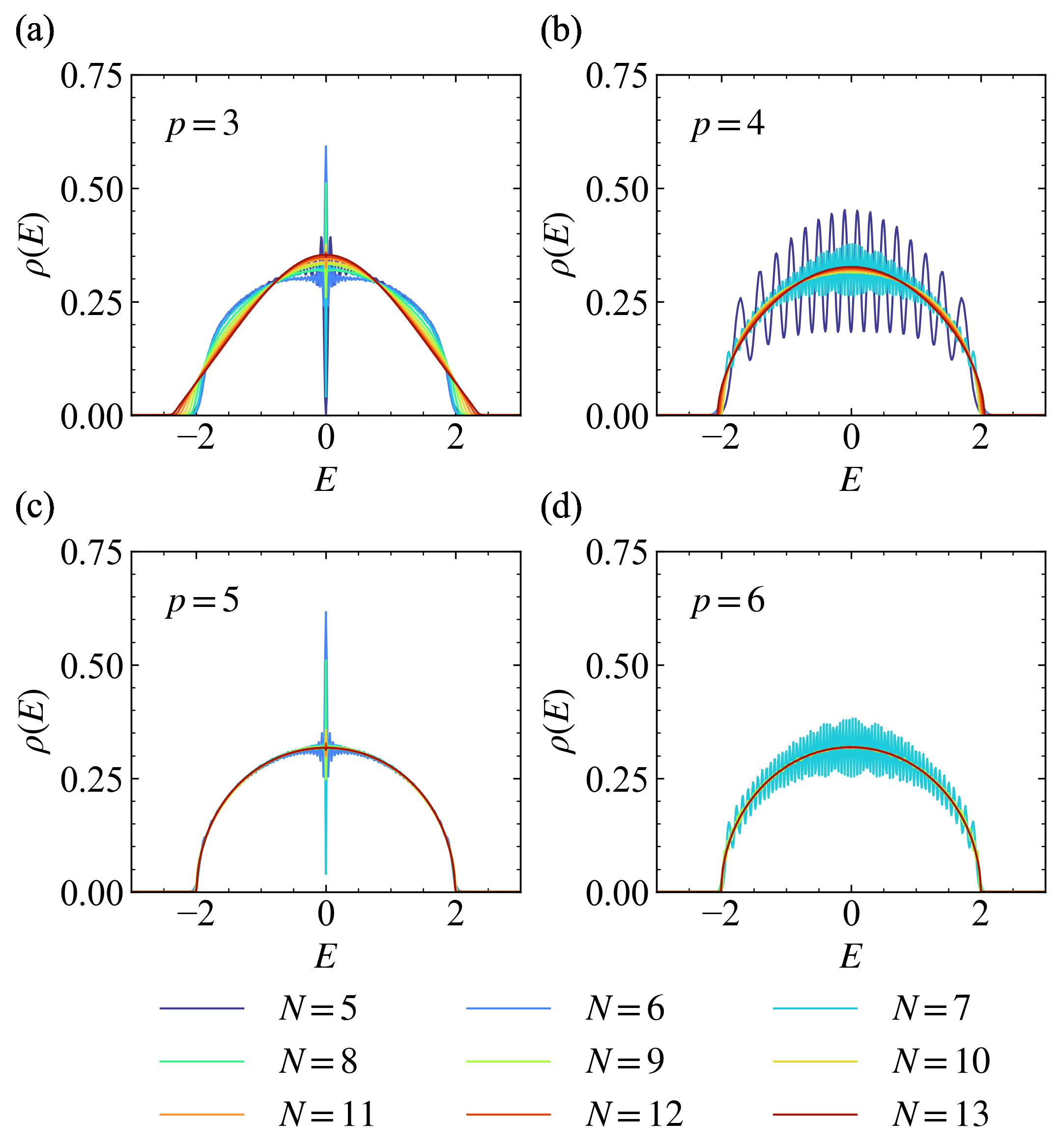}
\vspace{-20pt}
\caption{Density of states for 
the isotropic $p$-spin model with \((\varepsilon_x, \varepsilon_y) = (1, 1)\) for (a) \( p = 3 \), (b) \( p = 4 \), (c) \( p = 5 \), and (d) \( p = 6 \). The data are obtained by averaging over \( 2^{24 - N} \) random samples.}
\label{fig:dos_isotropic}
\end{figure}

Correspondingly, we find a characteristic change in the density of states \( \rho(E) \) shown in Fig.~\ref{fig:dos_isotropic}. \( \rho(E) \) is known to exhibit a crossover from the Wigner semicircle distribution to the normal distribution associated with a quantum-classical transition~\cite{Erdos2014}. As shown in Fig.~\ref{fig:dos_isotropic}(a) for \( p = 3 \), such a crossover is observed with increasing \(N\), except for the sharp feature around the zero energy for small $N$ that was also observed in Ref.~\cite{Erdos2014}. The behavior appears consistent with the prediction \( p \sim N^{1/2}\)~\cite{Erdos2014}, where \(p=3\) gives \(N \sim 9\). This crossover scaling agrees with the change of \( q_{\mathrm{EA}} \) in Fig.~\ref{fig:qEA_isotropic}(a), which indicates the transition between spin-liquid and spin-glass behavior. In contrast, for \( p = 4 \) in Fig.~\ref{fig:dos_isotropic}(b), while a slight indication of a crossover is seen, \( \rho(E) \) closely resembles the Wigner semicircle form except for the oscillation for small $N$~\cite{Erdos2014} within the system sizes shown here. For \( p = 5 \) and \( 6 \), \( \rho(E) \) obeys the Wigner semicircle form, and no such crossover is visible. These results are consistent with the changes of \( q_{\mathrm{EA}} \) with $N$, supporting the above classification of the spin-liquid and spin-glass states based on the decreasing or increasing trend of \( q_{\mathrm{EA}} \).

\subsection{Spin liquid under more restricted conditions in the $XY$ limit\label{sec:XY}}
Next, we show the results for the $XY$ limit, the $XY$ $p$-spin model with \((\varepsilon_x, \varepsilon_y) = (1, 0)\) and \(h = 0\) in Eq.~\eqref{eq:H_general}. We elucidate how the transition between spin liquid and spin glass, observed in the previous section for the isotropic case, is altered in the $XY$ limit, by clarifying the conditions under which the spin liquid can exist.

\begin{figure}[tb]
\centering
\includegraphics[width=\columnwidth]{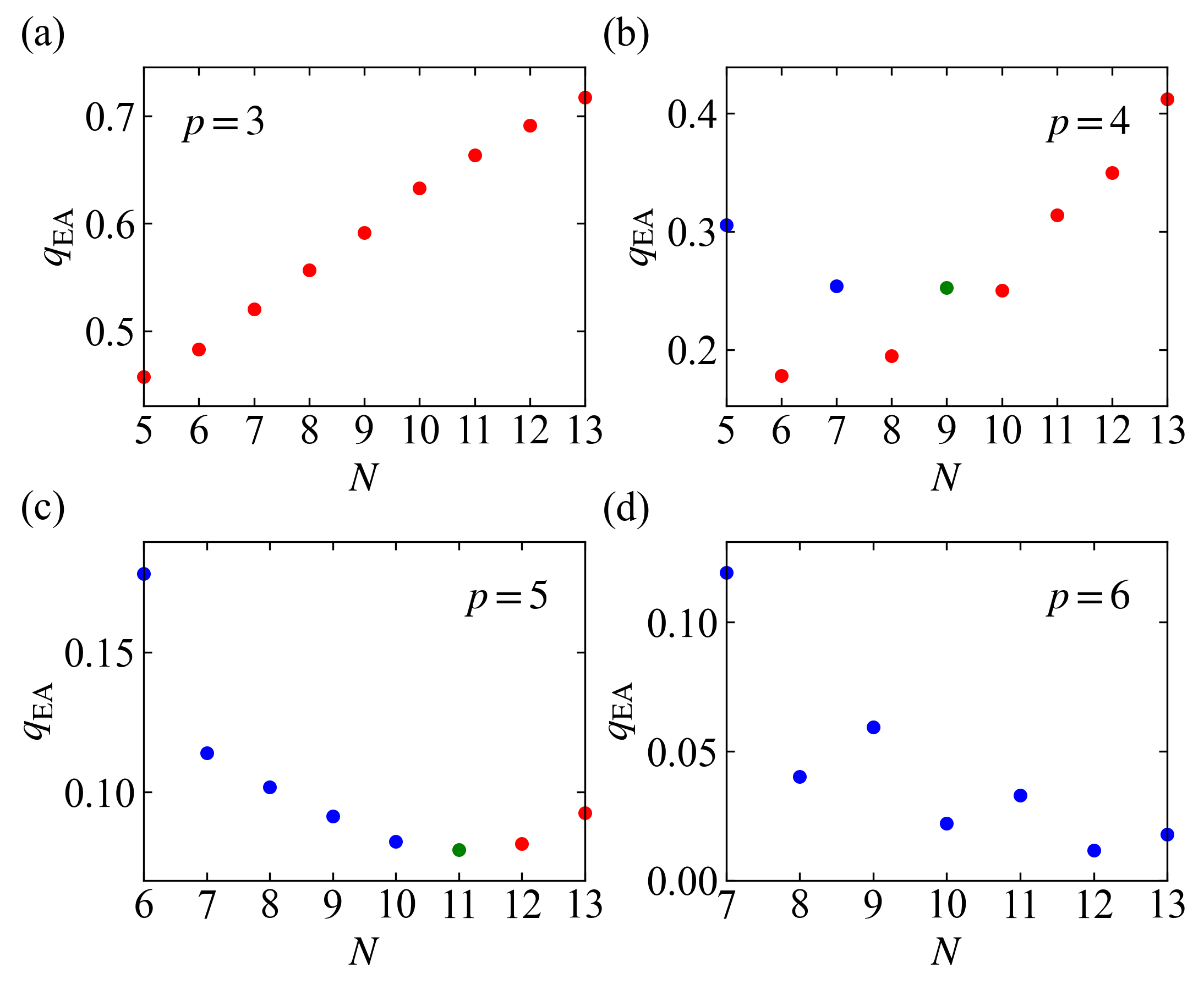}
\vspace{-20pt}
\caption{Corresponding plots to Fig.~\ref{fig:qEA_isotropic} for the $XY$ $p$-spin model with $(\varepsilon_x, \varepsilon_y)=(1, 0)$. The data are obtained by averaging over 300 random samples.}
\label{fig:qEA_XY}
\end{figure}

Similar to Sec.~\ref{sec:isotropic}, we begin by examining \( q_{\mathrm{EA}} \). The $N$ dependence of \( q_{\mathrm{EA}} \) for $p = 3$, $4$, $5$, and $6$ is shown in Fig.~\ref{fig:qEA_XY}. In contrast to the isotropic case in Fig.~\ref{fig:qEA_isotropic}, \( q_{\mathrm{EA}} \) monotonically increases with $N$ for \( p=3 \). Instead, a change from a decreasing to an increasing trend, which was observed for \( p = 3 \) in Fig.~\ref{fig:qEA_isotropic}, is now seen for \( p = 5 \) at $N=11$. For even \( p \), an oscillation between even and odd $N$ is still present as observed in Fig.~\ref{fig:qEA_isotropic}. However, for the $p=4$ case, while \( q_{\mathrm{EA}} \) increases monotonically for even $N$, it exhibits a similar change from a decreasing to an increasing trend at $N=9$ for odd $N$. For \( p = 6 \), \( q_{\mathrm{EA}} \) shows only a decreasing trend, except for the oscillating behavior. Following the criterion in Sec.~\ref{sec:isotropic}, we conclude that these data suggest the conditions $N$ and $p$ for the transition between the spin-liquid and spin-glass states are altered in the current $XY$ limit; the region of the spin-glass state expands from the isotropic case and is observed for \(p = 3\), $4$, and $5$, and accordingly, the region of the spin-liquid state is more restricted than in the isotropic case.

\begin{figure}[tb]
\centering
\includegraphics[width=\columnwidth]{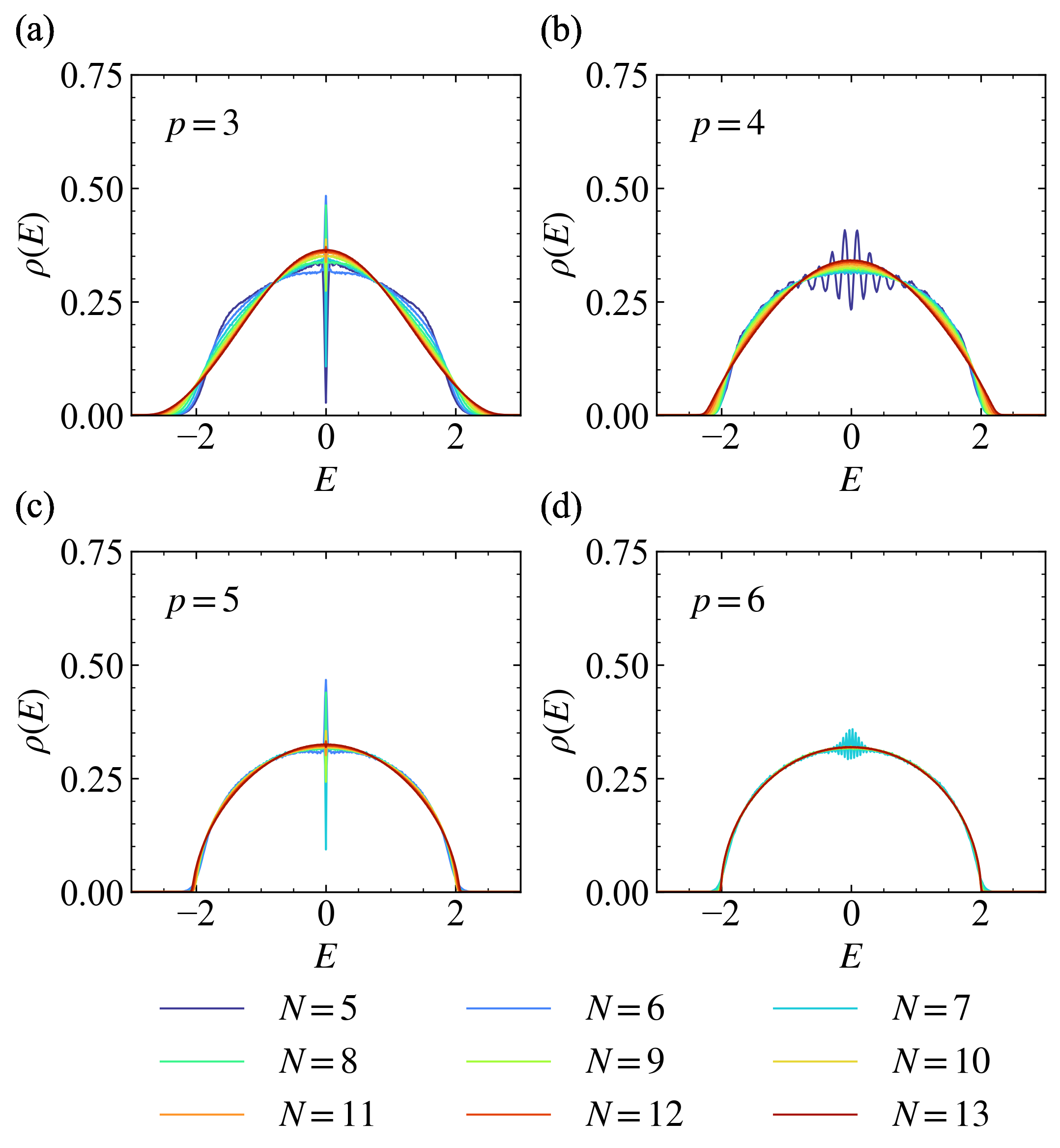}
\vspace{-20pt}
\caption{Corresponding plots to Fig.~\ref{fig:dos_isotropic} for the $XY$ $p$-spin model with $(\varepsilon_x, \varepsilon_y)=(1, 0)$. The data are obtained by averaging over \( 2^{24 - N} \) random samples.}
\label{fig:dos_XY}
\end{figure}

We also examine \( \rho(E) \) in Fig.~\ref{fig:dos_XY}. In contrast to Fig.~\ref{fig:dos_isotropic}, \( \rho(E) \) for \( p=3 \) behaves like a normal distribution already from the smallest \(N\) calculated. In addition, we observe a symptom of a crossover from the Wigner semicircle form to the normal one for \( p=4 \) and \(5\). This suggests that a larger value of $p$ is required for the crossover in \( \rho(E) \) than the isotropic case at the same $N$, as suggested by \( q_{\mathrm{EA}} \) above.

In the Appendix, we derive the general form of \( \rho(E) \) in the thermodynamic limit for the anisotropic case, indicating that the crossover in the current $XY$ case occurs at a $p$ value $2/\sqrt{3}$ times larger than that in the isotropic case. This aligns well with the trends found in \( q_{\mathrm{EA}} \) and \( \rho(E) \), where the boundary between the spin-liquid and spin-glass states shifts to larger $p$.

\subsection{Absence of spin liquid in the Ising limit\label{sec:Ising}}
\begin{figure}[tb]
\centering
\includegraphics[width=\columnwidth]{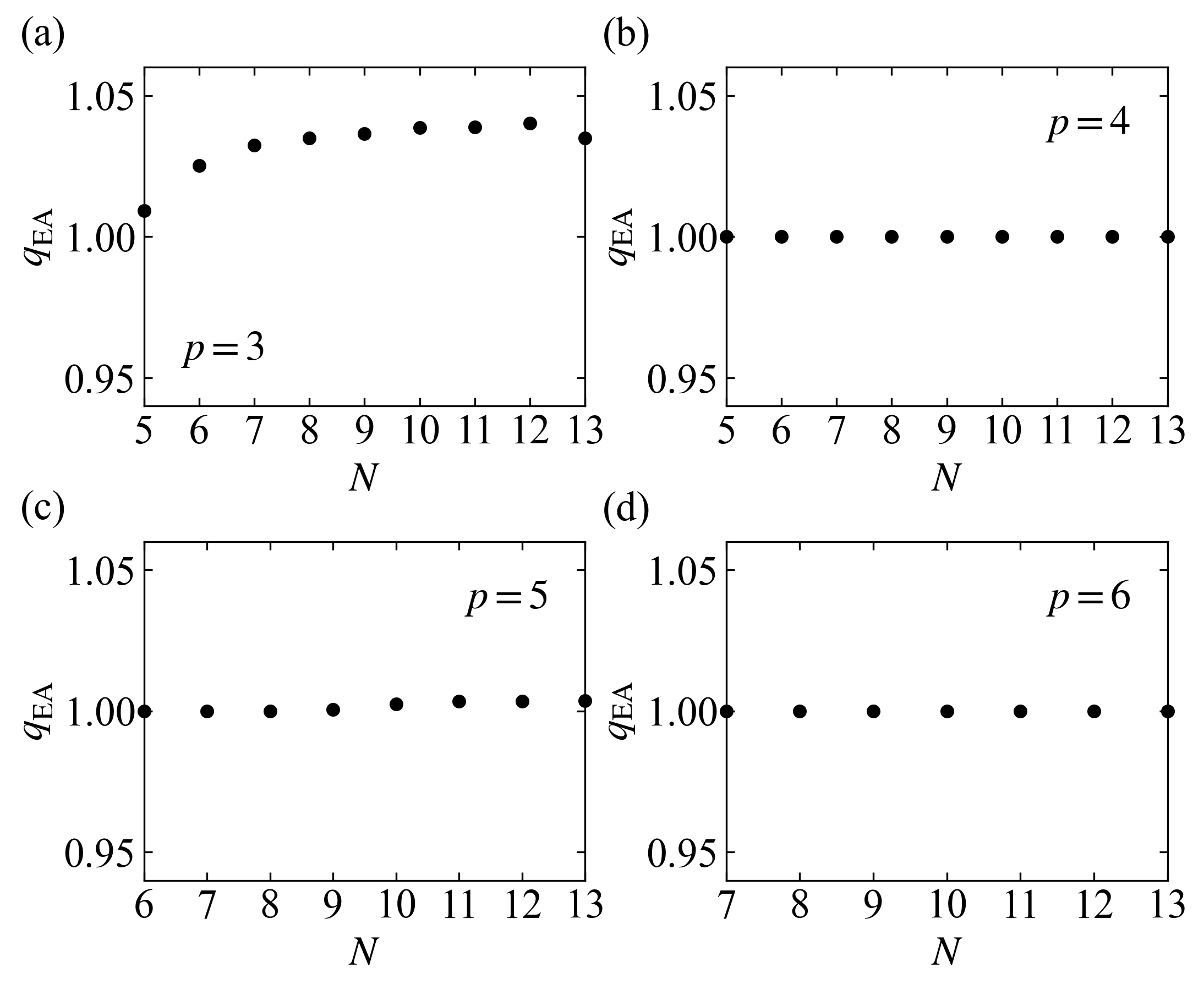}
\vspace{-20pt}
\caption{Corresponding plots to Fig.~\ref{fig:qEA_isotropic} for the classical $p$-spin model with $(\varepsilon_x, \varepsilon_y)=(0, 0)$. All data are plotted as black dots due to the absence of a clearly discernible increasing or decreasing trend. The data are obtained by averaging over 300 random samples.}
\label{fig:qEA_Ising}
\end{figure}

\begin{figure}[tb]
\centering
\includegraphics[width=\columnwidth]{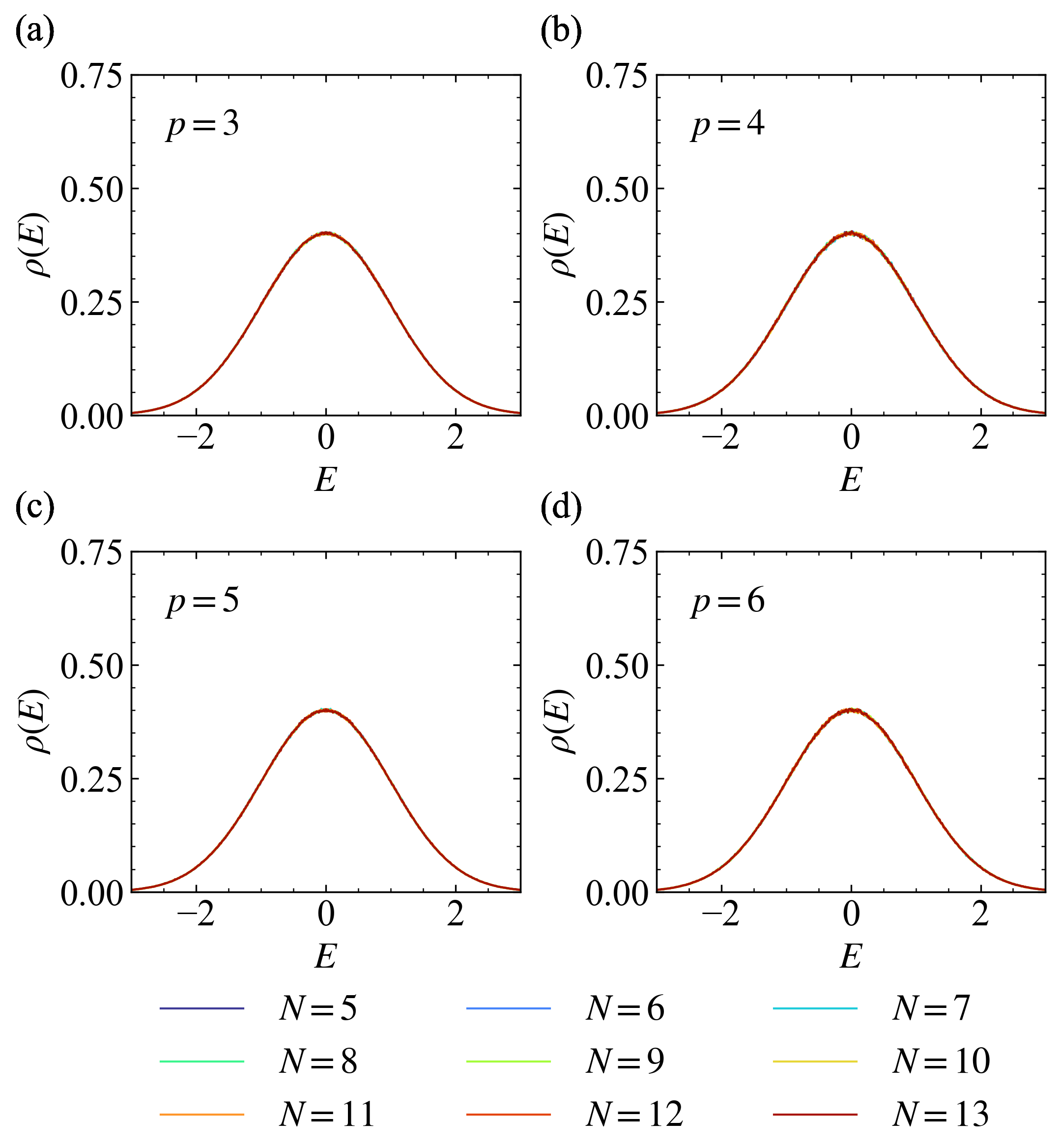}
\vspace{-20pt}
\caption{Corresponding plots to Fig.~\ref{fig:dos_isotropic} for the classical $p$-spin model with $(\varepsilon_x, \varepsilon_y)=(0, 0)$. The data are obtained by averaging over \( 2^{24 - N} \) random samples.}
\label{fig:dos_Ising}
\end{figure}

Finally, we show the results for the classical $p$-spin model with Ising-type interactions, where the parameters of the model in Eq.~\eqref{eq:H_general} are set to \((\varepsilon_x, \varepsilon_y) = (0, 0)\) and \(h = 0\). Figure~\ref{fig:qEA_Ising} displays the $N$ dependence of \( q_{\mathrm{EA}} \). We find that \( q_{\mathrm{EA}} \) is close to $1$ with less dependence on $N$ for all $p$. Note that \( q_{\mathrm{EA}} \) exceeding $1$ is due to the definition in Eq.~\eqref{eq:qEA} using the two lowest-energy states. The results indicate that the system shows robust spin-glass behavior in this Ising limit, regardless of $N$ and $p$. Similar behavior was observed for the limited cases in the previous studies~\cite{Derrida1980, Gardner1985, Kirkpatrick1987}. Correspondingly, the density of states takes the form of a normal distribution for all combinations of \(N\) and \(p\), as shown in Fig.~\ref{fig:dos_Ising}.

\subsection{Phase diagram\label{sec:diagram}}
\begin{figure}[tbp]
\centering
\includegraphics[width=\columnwidth]{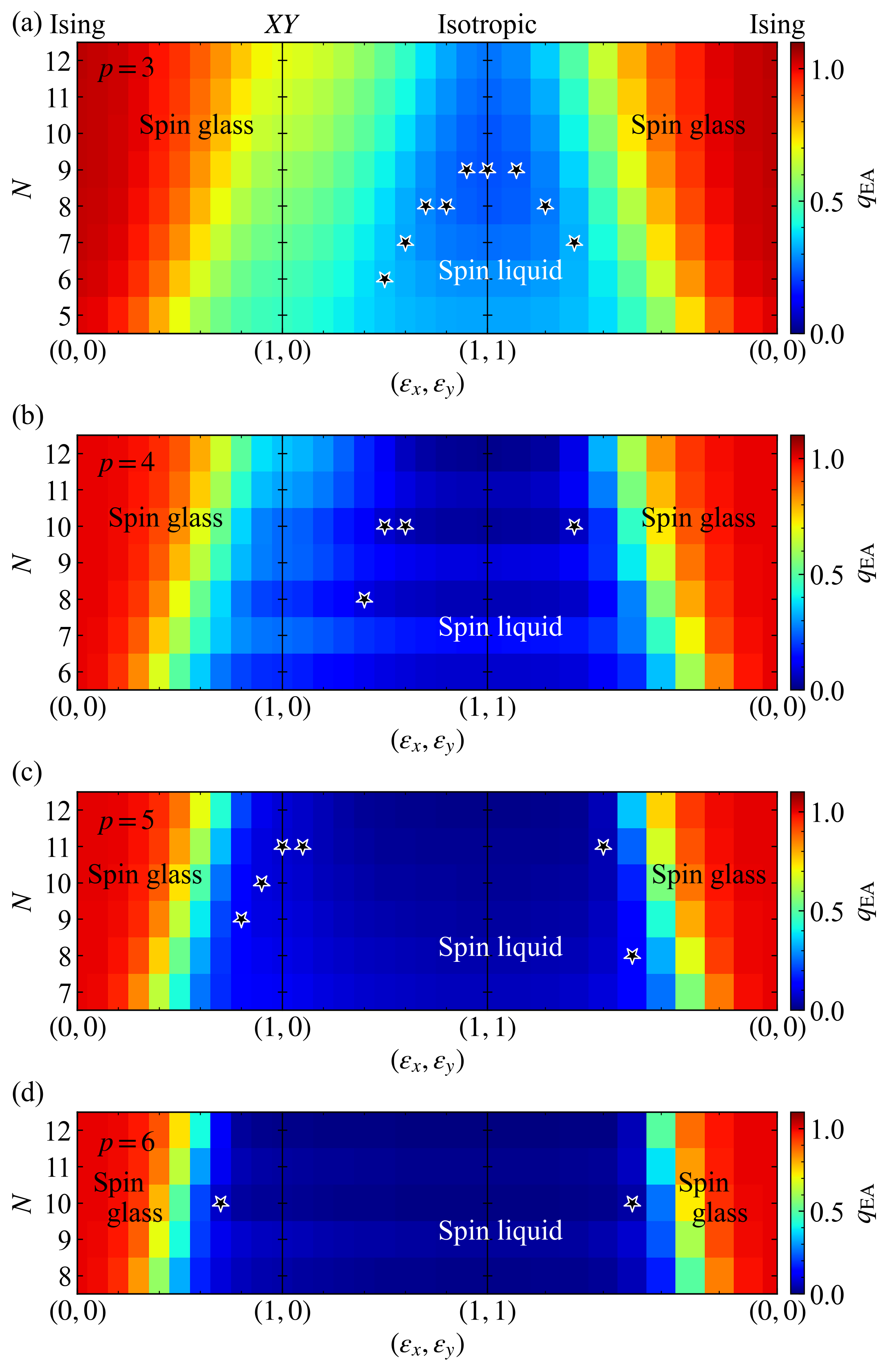}
\caption{Phase diagrams by the EA order parameter \( q_{\mathrm{EA}} \) for (a) \( p = 3 \), (b) \( p = 4 \), (c) \( p = 5 \), and (d) \( p = 6 \). The color represents the values of \( q_{\mathrm{EA}} \), and the star symbols indicate the system sizes \( N \) minimizing \( q_{\mathrm{EA}} \) for each \((\varepsilon_x, \varepsilon_y)\). The symbols are displayed only if the corresponding $N$ is smaller than $12$.  The data are obtained by averaging over 500 random samples.}
\label{fig:diagram3-6}
\end{figure}

Here, we study the phase diagram by connecting the three points \((\varepsilon_x, \varepsilon_y) = (1, 1)\), \((1, 0)\), and \((0, 0)\) studied in the previous sections. As a path to connect these points, the following ranges of parameters \((\varepsilon_x, \varepsilon_y)\) are chosen (see also Fig.~\ref{fig:epsilon}). (i) From Ising to $XY$ by varying \(\varepsilon_x\) from 0 to 1 with \(\varepsilon_y = 0\): In this case, the spin components included in the Hamiltonian Eq.~\eqref{eq:H_general} change from \(\{z\}\) to \(\{z, x\}\), gradually incorporating the \(x\) component into the interactions. (ii) From $XY$ to isotropic by varying \(\varepsilon_y\) from 0 to 1 with \(\varepsilon_x = 1\): The spin components change from \(\{z, x\}\) to \(\{z, x, y\}\), gradually incorporating the \(y\) component into the interactions. (iii) From isotropic to Ising by varying both \(\varepsilon_x\) and \(\varepsilon_y\) from 1 to 0 simultaneously with \(\varepsilon_x = \varepsilon_y\): This path reduces the spin components from \(\{z, x, y\}\) to \(\{z\}\), gradually removing both \(x\) and \(y\) components from the interactions. This approach allows us to construct the phase diagrams shown in Fig.~\ref{fig:diagram3-6}, which are given by a color map of \( q_{\mathrm{EA}} \) for the case of \( p = 3, 4, 5\), and $6$, with \( N \) ranging from \( p+2 \) to 12. The vertical cuts at $(\varepsilon_x, \varepsilon_y) = (1, 1)$, $(1, 0)$, and $(0, 0)$ in Fig.~\ref{fig:diagram3-6} correspond to Figs.~\ref{fig:qEA_isotropic}, \ref{fig:qEA_XY}, and \ref{fig:qEA_Ising}, respectively. In Fig.~\ref{fig:diagram3-6}, calculated points along the horizontal axis are taken for every $0.1$ increase or decrease in \(\varepsilon_x\), \(\varepsilon_y\), or both, starting from \((\varepsilon_x, \varepsilon_y) = (0, 0)\). The star symbols indicate the system sizes $N$ minimizing \( q_{\mathrm{EA}} \) for each \((\varepsilon_x, \varepsilon_y)\). These symbols are omitted if given by \( N=12 \) which is the largest value of \( N \) calculated, as \( q_{\mathrm{EA}} \) may continue to decrease for larger \( N \).

The phase diagrams in Fig.~\ref{fig:diagram3-6} include the spin-liquid and spin-glass states assigned in Figs.~\ref{fig:qEA_isotropic}, \ref{fig:qEA_XY}, and \ref{fig:qEA_Ising}. For \( p = 3 \), at small \( N \) around \((\varepsilon_x, \varepsilon_y) = (1, 1)\), \( q_{\mathrm{EA}} \) decreases with \(N\), indicating spin-liquid behavior. Outside of this range, i.e., when \((\varepsilon_x, \varepsilon_y)\) moves away from \((1, 1)\) or when \(N\) increases, the system exhibits the spin-glass behavior where \(q_{\mathrm{EA}}\) increases with \(N\). Thus, the boundary indicated by the star symbols becomes a domelike shape for the spin-liquid state in the phase diagram. For \( p = 4 \), a similar trend is observed, but the spin-liquid region becomes larger, and the symbols around the isotropic case are not numerically identified within the range of $N$ calculated. For \( p = 5 \) and \( p = 6 \), the spin-liquid region also appears to expand, and the spin-liquid behavior is observed even near the $XY$ case at small \( N \).

\begin{figure}[tbp]
\centering
\includegraphics[width=\columnwidth]{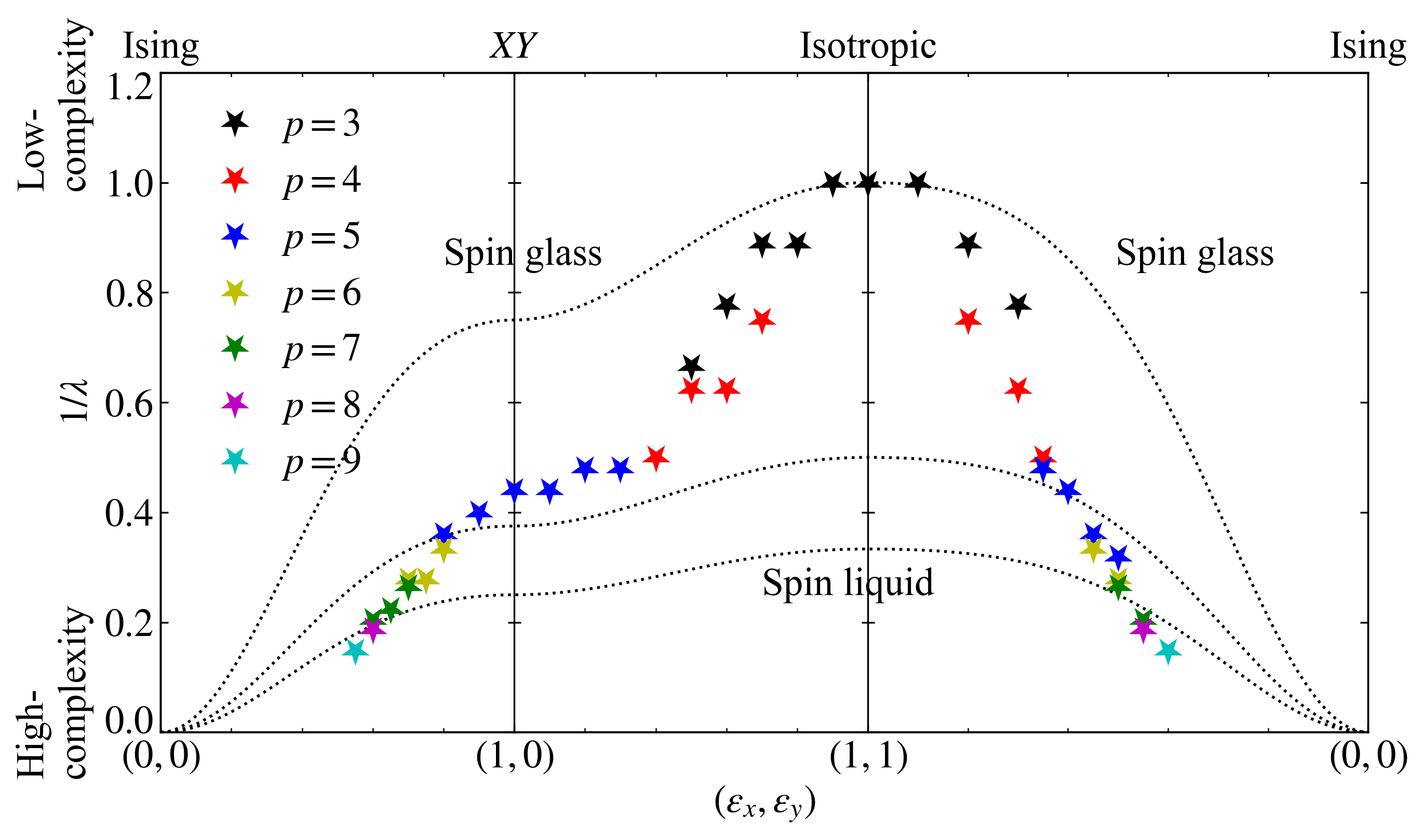}
\caption{Phase diagram summarized by the value of $1/\lambda = N/p^2$. The star symbols indicate the system sizes $N$ minimizing the EA order parameter \( q_{\mathrm{EA}} \) for \(p = 3, 4, 5, 6, 7, 8\), and $9$, within the plot range for each \((\varepsilon_x, \varepsilon_y)\). The data are obtained by averaging over 100 random samples. The dotted lines represent the contours for $\tilde{\lambda} = 1, 2$, and $3$ from top to bottom, where \(\tilde{\lambda}\) is given in Eq.~\eqref{eq:lambda_tilde} obtained by taking the thermodynamic limit with $\lambda$ fixed.}
\label{fig:diagram3-9}
\end{figure}

The trend in Fig.~\ref{fig:diagram3-6} and the behavior of the density of states in Figs.~\ref{fig:dos_isotropic}, \ref{fig:dos_XY}, and \ref{fig:dos_Ising} motivate us to summarize the phase diagram with respect to \( \lambda=p^2/N \). Figure~\ref{fig:diagram3-9} plots the system sizes $N$ minimizing \(q_{\mathrm{EA}}\) for all $p$ with respect to $1/\lambda = N/p^2$. Note that the accessible range of $1/\lambda$ is limited for each value of $p$, as the available system size $N$ is limited for each $p$. Surprisingly, as shown in Fig.~\ref{fig:diagram3-9}, all the data appear to collapse onto a single curve, suggesting that the phase diagrams for different $p$ and $N$ can be summarized in a unified manner by the single parameter $1/\lambda$. This also implies that the model in Eq.~\eqref{eq:H_general} exhibits a phase transition between the spin-liquid and spin-glass phases in the thermodynamic limit of $N\to \infty$ when $p$ is taken to be infinity with keeping $1/\lambda$ a constant, namely, a simultaneous infinity limit of $p$ and $N$ with \( p^2/N \) fixed. Our current numerical data are insufficient to analyze the critical behavior of this transition; determining the critical point and scaling behavior is left for future study.

We compare the plots in Fig.~\ref{fig:diagram3-9} obtained from \(q_{\mathrm{EA}}\) with the crossover in the density of states. 
The dotted lines in Fig.~\ref{fig:diagram3-9} represent the contours of \( \tilde{\lambda} \), which is a parameter characterizing the crossover of the density of states in the thermodynamic limit given by
\begin{align}
\tilde{\lambda} = \frac{3(\varepsilon_x^2 + \varepsilon_y^2 + \varepsilon_x^2 \varepsilon_y^2)}{(1 + \varepsilon_x^2 + \varepsilon_y^2)^2}\lambda.
\label{eq:lambda_tilde}
\end{align}
This is obtained by generalizing the density of states for the isotropic case derived in Ref.~\cite{Erdos2014} to the model in Eq.~\eqref{eq:H_general}, where anisotropy is introduced through \((\varepsilon_x, \varepsilon_y)\) (see the Appendix). By comparison, we find a similar trend between the contours of \( \tilde{\lambda} \) and the phase boundary obtained from \( q_{\mathrm{EA}} \). Both have maxima in the isotropic case and gradually decrease toward the Ising limit with a plateau around the $XY$ limit. The phase boundary by \( q_{\mathrm{EA}} \) does not fall onto a single contour of \( \tilde{\lambda} \). We note that a similar trend was already seen in Secs.~\ref{sec:isotropic} and \ref{sec:XY}. The crossover observed in Fig.~\ref{fig:dos_XY}(c) is less pronounced than that in Fig.~\ref{fig:dos_isotropic}(a), while the corresponding \( q_{\mathrm{EA}} \) exhibits a crossover in both cases, as shown in Figs.~\ref{fig:qEA_XY}(c) and \ref{fig:qEA_isotropic}(a).

\subsection{Competition between multiple-spin interactions and an external magnetic field\label{sec:field}}
Finally, by applying an external magnetic field \( h \) to the isotropic case with \( (\varepsilon_x, \varepsilon_y) = (1, 1) \), we investigate how the induced anisotropy competes with multiple-spin interactions and alters the ground-state properties. In the limit of $h\to \infty$, the system is in a trivial quantum paramagnetic state (forced ferromagnetic state). We investigate how the spin liquid and the spin glass at $h=0$ evolve into the quantum paramagnetic state as $h$ increases. To identify such transitions, we introduce the EA order parameter calculated from the \( x \) and \( y \) components of the spins, explicitly excluding the \( z \) component that aligns with the external field:
\begin{align}
q^{xy}_{\mathrm{EA}} = \frac{1}{2N} \sum_{i} \sum_{\alpha \in \{x, y\}} \sum_{a,b=1,2} \left| \langle \psi_a | \sigma_i^\alpha | \psi_b \rangle \right|^2.
\label{eq:qxyEA}
\end{align}
In the isotropic case with \( h = 0 \), this quantity behaves qualitatively the same as \( q_{\mathrm{EA}} \) given in Eq.~\eqref{eq:qEA}.

\begin{figure}[tb]
\centering
\includegraphics[width=\columnwidth]{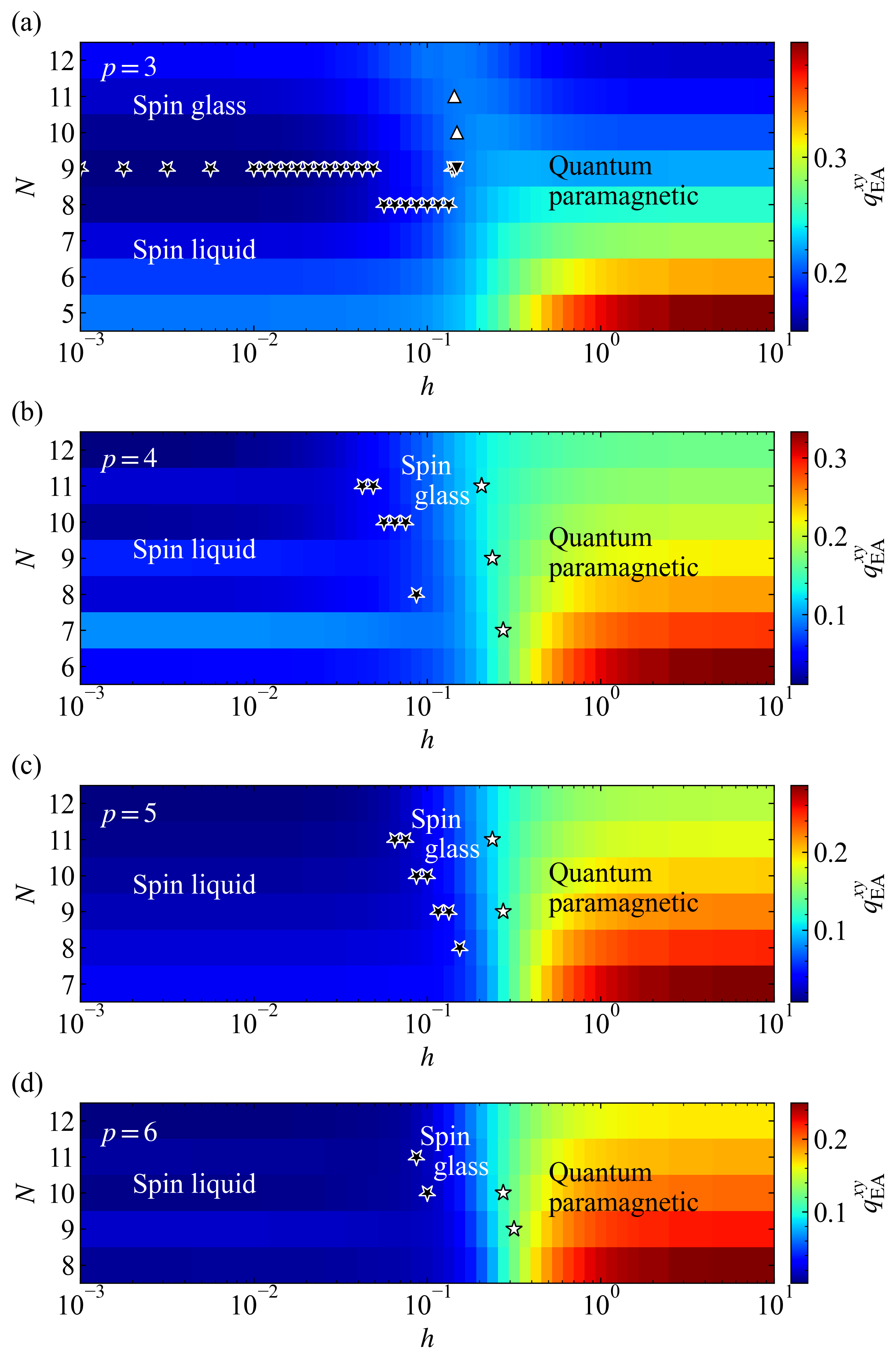}
\caption{Phase diagrams by the EA order parameter defined by the $x$ and $y$ components, $q^{xy}_{\mathrm{EA}}$, for the isotropic $p$-spin model with (a) \( p = 3 \), (b) \( p = 4 \), (c) \( p = 5 \), and (d) \( p = 6 \) under the external magnetic field $h$ in the $z$ direction. The color represents the values of \( q^{xy}_{\mathrm{EA}} \). The black and white star symbols represent the system sizes \( N \) minimizing \( q^{xy}_{\mathrm{EA}} \) and \( N \) maximizing \( q^{xy}_{\mathrm{EA}} \), respectively, for each $h$. In (a), the black and white triangular symbols represent \( N \) locally minimizing \( q^{xy}_{\mathrm{EA}} \) and  \( N \) locally maximizing \( q^{xy}_{\mathrm{EA}} \), respectively. The symbols are displayed only if the corresponding $N$ is smaller than $12$, and the white star symbols are omitted for $h \leq 0.15$. The data are obtained by averaging over 100 random samples.}
\label{fig:qxyEA_p=3-6}
\end{figure}

Figure~\ref{fig:qxyEA_p=3-6} displays the $h$ and $N$ dependence of $q^{xy}_{\mathrm{EA}}$ for different values of $p$. In this figure, black and white star symbols indicate the system sizes \( N \) minimizing \( q^{xy}_{\mathrm{EA}} \) and \( N \) maximizing \( q^{xy}_{\mathrm{EA}} \), respectively, for each $h$. Only for $p = 3$, black and white triangular symbols are also used to indicate \( N \) locally minimizing \( q^{xy}_{\mathrm{EA}} \) and  \( N \) locally maximizing \( q^{xy}_{\mathrm{EA}} \), respectively. As in Fig.~\ref{fig:diagram3-6}, the black symbols are used to identify the boundary between the spin-liquid and spin-glass phases. On the other hand, the white symbols are used to capture the transition to the paramagnetic phase.

We first discuss the case of \( p = 3 \), shown in Fig.~\ref{fig:qxyEA_p=3-6}(a). For low $h$ $\lesssim 0.5$, we observe the same behavior as in the zero-field case in Fig.~\ref{fig:diagram3-6}(a) with \((\varepsilon_x, \varepsilon_y) = (1, 1)\): $q^{xy}_{\mathrm{EA}}$ initially decreases as $N$ increases, but then begins to increase for $N>9$, indicating a transition from the spin liquid to the spin glass at $N\simeq 9$. As $h$ increases, the minimum shifts from $N = 9$ to $8$, and finally, the spin-glass-like increase of \( q^{xy}_{\mathrm{EA}} \) disappears for $h \gtrsim 0.15$, where $q^{xy}_{\mathrm{EA}}$ decreases monotonically for all $N$. This suggests that the spin-glass state becomes unstable and changes into a quantum paramagnetic state.

Next, we examine the case of $p = 4$, shown in Fig.~\ref{fig:qxyEA_p=3-6}(b). In this case, for low $h$, the minimum of $q^{xy}_{\mathrm{EA}}$ is out of range of $N$ calculated, as shown in the zero-field case in Fig.~\ref{fig:qEA_isotropic}(b). However, for higher \( h \gtrsim 0.04\), a spin-glass-like increase becomes visible within the range, and the minimum shifts to smaller $N$. This shift is more pronounced than in the \( p = 3 \) case and reaches the smallest $N$ at \( h \sim 0.1 \). At $h \sim 0.2$, $q^{xy}_{\mathrm{EA}}$ exhibits a maximum, suggesting a transition from the spin glass to the paramagnetic state. This indicates that in contrast to the $p=3$ case, the spin-glass state intervenes between the spin-liquid and paramagnetic states; the system exhibits successive transitions from the spin liquid to the spin glass, and then to the paramagnetic state as $h$ increases, regardless of the calculated value of \( N \). We observe similar behavior for \( p = 5 \) and \( p = 6 \), as shown in Figs.~\ref{fig:qxyEA_p=3-6}(c) and \ref{fig:qxyEA_p=3-6}(d), respectively. This suggests that the qualitatively different behavior in the $p=3$ case is likely due to the small value of $p$, and the results for larger $p$ represent the generic behavior in the current model.

\begin{figure}[tb]
\centering
\includegraphics[width=\columnwidth]{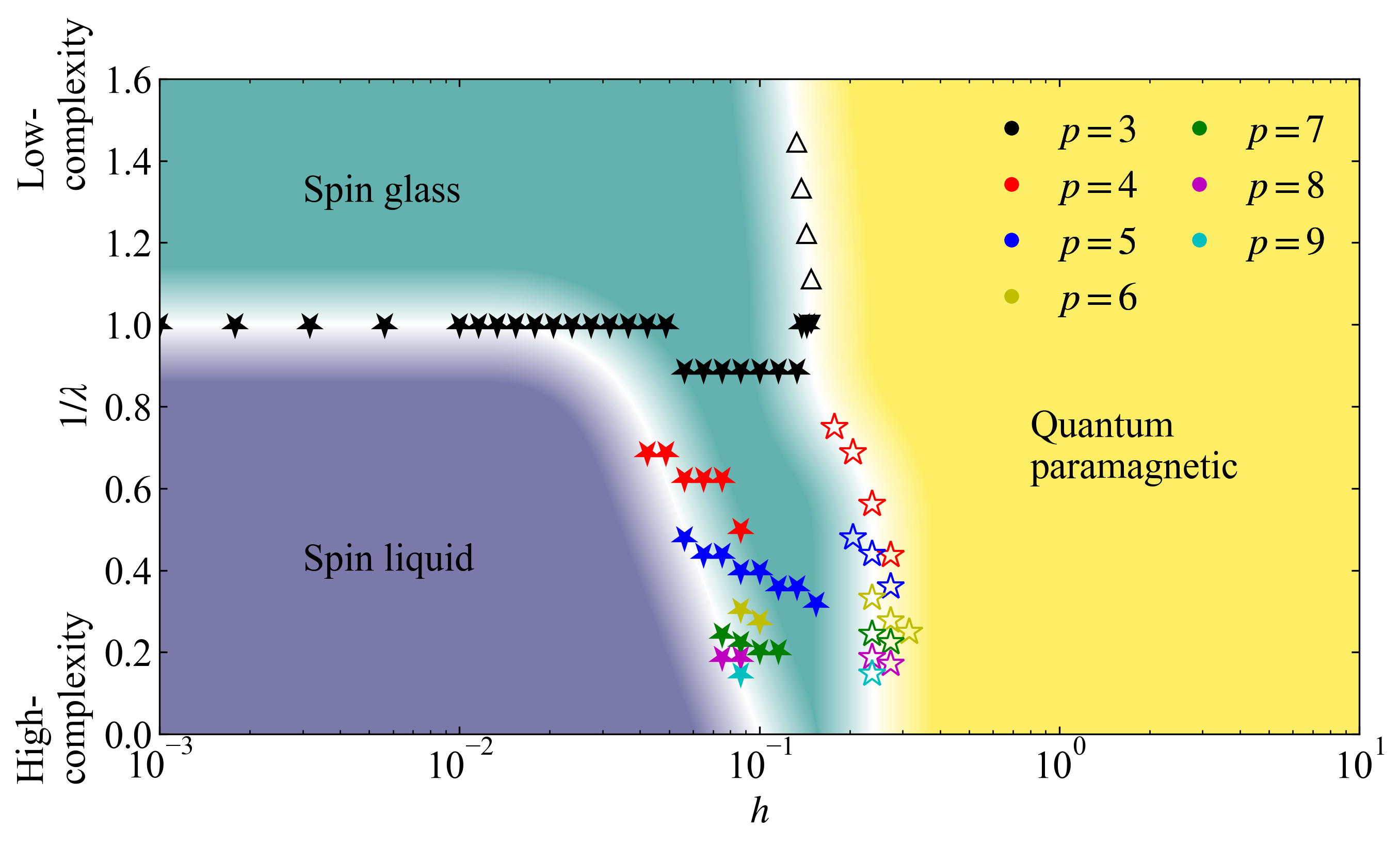}
\caption{
Phase diagram inferred by the value of $1/\lambda = N/p^2$. The symbols are common to those in Fig.~\ref{fig:qxyEA_p=3-6}, with each $p$ represented by the color indicated in the legend. The data are obtained by averaging over 100 random samples.
}
\label{fig:qxyEA_p=3-9}
\end{figure}

Following the approach in Fig.~\ref{fig:diagram3-9}, here we summarize the phase diagrams under the external magnetic field for multiple values of $p$ in Fig.~\ref{fig:qxyEA_p=3-9} sorted by $1/\lambda = N/p^2$. We find three distinct phases: the spin liquid in the region where $1/\lambda \lesssim 1$ and $h\lesssim 0.1$, the quantum paramagnetic phase for $h\gtrsim 0.3$, and the spin-glass phase extending from the region where $1/\lambda \gtrsim 1$ and $h\lesssim 0.1$ to $1/\lambda \lesssim 1$ and $0.1\lesssim h\lesssim 0.3$. This result demonstrates that the spin-liquid phase arising from multiple-spin interactions in the model in Eq.~\eqref{eq:H_general} transitions into the spin-glass phase before entering the quantum paramagnetic phase. Given that introducing anisotropy in the interaction terms leads to a transition from the spin-liquid to the spin-glass phase (see Sec.~\ref{sec:diagram}), the anisotropy effectively induced by the external magnetic field is also expected to destabilize the spin liquid and facilitate the emergence of intervening spin-glass behavior. 

We note that the boundary between the spin-liquid and spin-glass states signaled by the minimum of $q^{xy}_{\mathrm{EA}}$ decreases as \( h \) increases, whereas it does not collapse onto a single curve. This is also the case for the boundary to the paramagnetic state signaled by the maxima of \( q^{xy}_{\mathrm{EA}} \), while the deviations are smaller. These are possibly due to finite-size effects. Further studies are needed to clarify the behavior in the thermodynamic limit.

\section{Concluding remarks\label{sec:conclusion}}
To summarize, we have investigated the transition between two representative quantum disordered states, spin-liquid and spin-glass states, by studying a quantum spin model with all-to-all random $p$-spin interactions, inspired by extensions of the Sachdev-Ye-Kitaev model. In addition to varying $p$, we introduced anisotropy parameters for the interactions, $(\varepsilon_x, \varepsilon_y)$, which allow for a smooth interpolation between the Ising-like one-component, $XY$-like two-component, and isotropic three-component spin models. By computing the Edwards-Anderson order parameter and the density of states, we revealed a crossover between the spin-liquid and spin-glass states, with the crossover point depending sensitively on $p$, $(\varepsilon_x, \varepsilon_y)$, and the system size $N$. We elucidated the phase diagram through scaling with $N/p^2$, suggesting a phase transition between these two states in the thermodynamic limit where $N/p^2$ remains constant. We showed that the phase boundary is sensitive to the interaction anisotropy; the spin-liquid phase is most prominent in the isotropic case, rapidly diminishes with increasing anisotropy, and ultimately disappears in the Ising limit. The tendency that spin-liquid behavior becomes more pronounced with increasing interaction complexity (larger $p^2/N$) and decreasing anisotropy in spin space $[(\varepsilon_x, \varepsilon_y) \to (1,1)]$ can be understood as a consequence of enhanced frustration and quantum fluctuations.

We also examined the transition from the spin liquid to a quantum paramagnetic state by applying an external magnetic field to the model with isotropic interactions. Our results showed that as the field strength increases, the spin-liquid phase transitions into the spin-glass phase and eventually enters a quantum paramagnetic phase. It is likely that the anisotropy induced by the external magnetic field, similarly to that introduced in the interaction terms, also contributes to the destabilization of the spin liquid and the emergence of the intervening glass behavior.

Our results show that the transition from spin-liquid to spin-glass states is strongly influenced by the complexity of random all-to-all interactions, the anisotropy in spin space, and the external magnetic field, all of which govern magnetic frustration and quantum fluctuations. Furthermore, our systematic study uncovers the characteristic scaling of the phase boundary with $N/p^2$, in agreement with previous studies for several specific cases~\cite{Erdos2014, Berkooz2018, Berkooz2019}. These findings highlight that the quantum $p$-spin models offer a valuable platform for exploring the rich physics associated with spin-glass transitions.

An important direction for future work is to further extend the model to deepen our understanding of spin-liquid and spin-glass states. A straightforward extension involves introducing a sparse model, which is constructed as a sum of terms randomly sampled from the original all-to-all Hamiltonian. This approach allows more detailed analyses of the phase diagram and its scaling behavior with $N/p^2$. In the context of the Sachdev-Ye-Kitaev model, it was argued that sparse models recover the essential physics of the original formulation~\cite{Xu2020}. However, it remains unclear whether sparse models in spin systems can reproduce the properties of the original Hamiltonian. Addressing this question is critical for understanding the extent to which a reduction in the connectivity of interactions affects the behavior of quantum spin systems.

Another intriguing extension is to explore lattice models with finite-range interactions~\cite{Guo1994}. Such models are more realistic in the context of condensed matter physics and would provide valuable insights into the behavior of quantum magnets. In such finite-range models, each $p$-spin interaction involves spins located within a spatial region of size $L$ on a $d$-dimensional lattice. While the scaling $N/p^2$ in the all-to-all model suggests a phase boundary that scales with $N/L^{2d}$, the simplified interaction structure of finite-range models may significantly influence the presence or nature of this transition. To bridge the gap between the two regimes, a careful interpolation from the all-to-all model to the finite-range model, mediated by sparse models, is required.

In this work, we have primarily characterized the spin-liquid and spin-glass phases using the Edwards-Anderson order parameter. To further distinguish these phases, it will be important to investigate quantum entanglement properties, such as entanglement entropy. Moreover, since the distinction between these phases may also appear in dynamical quantities, one might expect corresponding signatures in the energy spectrum. Evaluating the level spacing statistics of the Hamiltonian, we found that Poisson statistics appear only in the classical Ising limit ($\varepsilon_x = \varepsilon_y = 0$), whereas in all other cases, the statistics follow the Gaussian ensemble (not shown). To gain further insight, it would also be interesting to calculate other indicators of quantum chaos. For example, in spin systems, analyzing the out-of-time-order correlator is essential for probing the glassy nature of the system and chaotic behavior. For a model with two-body random Pauli spin interactions, the out-of-time-order correlator can be computed~\cite{Zhou2019}, offering a promising avenue for further exploration.

\section*{Acknowledgments}
The authors thank K. Kobayashi for constructive suggestions. S.W. thanks S. Okumura for valuable discussions. This work was supported by the JSPS KAKENHI (Grants No. JP20H00122 and No. JP25H01247).

\appendix

\section{Derivation of the density of states in the thermodynamic limit\label{sec:A-dos}}
In this Appendix, we generalize the density of states in the thermodynamic limit of \( N \to \infty \) in the isotropic case~\cite{Erdos2014} to anisotropic cases with general values of \((\varepsilon_x, \varepsilon_y)\). In the isotropic case, the density of states was derived for finite \( \lambda_\infty = \lim_{N \to \infty} p^2/N \) as
\begin{align}
\rho_{\lambda_\infty}(E) &= \dfrac{\sqrt{1 - q}}{\pi \sqrt{1 - (1 - q)E^2/4}} \nonumber \\
&\quad \times \prod_{k=0}^\infty \left[ \dfrac{1 - q^{2k+2}}{1 - q^{2k+1}} \left( 1 - \dfrac{E^2(1 - q)q^k}{(1 + q^k)^2} \right) \right],
\label{eq:rho_lambda_infty}
\end{align}
where \( q = e^{-4\lambda_\infty/3} \)~\cite{Erdos2014}. This expression is valid for \( E \in \left[ -2/\sqrt{1 - q}, 2/\sqrt{1 - q} \right] \) and \( \rho_{\lambda_\infty}(E)=0 \) outside this range. To simplify the notation in Eq.~\eqref{eq:H_general}, we write \( I = (i_1, \ldots, i_p ) \) as the set of \( p \) distinct indices \( 1 \leq i_1 < \cdots < i_p \leq N \), and \( A = (\alpha_1, \ldots, \alpha_p) \) as the set of spin components \( \alpha_1, \ldots, \alpha_p \in \{x, y, z\} \). Using these definitions, the Hamiltonian can be expressed as
\begin{align}
H = (1 + \varepsilon_x^2 + \varepsilon_y^2)^{-p/2} \binom{N}{p}^{-1/2} \sum_{IA} J_{IA} \tilde{\sigma}_{I}^{A},
\label{eq:H_B}
\end{align}
where \( \tilde{\sigma}_{I}^{A} = \tilde{\sigma}_{i_1}^{\alpha_1} \cdots \tilde{\sigma}_{i_p}^{\alpha_p} \); \( \tilde{\sigma}^{\alpha} \) represents the scaled Pauli matrix defined in Eq.~\eqref{eq:sigma_tilde}.

Following the method in Ref.~\cite{Erdos2014}, we employ the moment method to compute the generalized density function.  
The $k$th moment is defined as
\begin{align}
m_k = \frac{1}{2^N} E \left[ \mathrm{Tr} \, H^k \right],
\end{align}
where $E[\cdots]$ denotes the disorder average.
In this model, all odd moments vanish, and the $k$th moment \( m_k \) for even moments can be expressed as
\begin{align}
m_k &= \lim_{N \to \infty} \frac{1}{(k/2)!} \sum_{\pi \in \mathcal{S}_k} (1 + \varepsilon_x^2 + \varepsilon_y^2)^{-kp/2} \binom{N}{p}^{-k/2} \nonumber \\
&\quad \times \sum_{I_1 A_1, \ldots, I_{k/2} A_{k/2}} 2^{-N} \mathrm{Tr} \tilde{\sigma}_{I_{\pi(1)}}^{A_{\pi(1)}} \cdots \tilde{\sigma}_{I_{\pi(k)}}^{A_{\pi(k)}},
\end{align}
where the set \( \mathcal{S}_k \) is defined as
\begin{align}
\mathcal{S}_k = \big\{ \pi : &\{1, \ldots, k\} \to \{1, \ldots, k/2\} \; \big| \nonumber \\
&\left| \pi^{-1}(\{j\}) \right| = 2 \; \text{for all} \; 1 \leq j \leq k/2 \big\},
\end{align}
which represents the set of pair partitions of \( k \) elements. Each partition \( \pi \in \mathcal{S}_k \) assigns every element of \( \{1, \ldots, k\} \) to one of the \( k/2 \) pairs, ensuring that exactly two elements are assigned to each pair. For a given partition \( \pi \in \mathcal{S}_k \), the number of crossings \( \kappa(\pi) \) is defined as the number of subsets \( \{r, s\} \subset \{1, \ldots, k/2\} \) such that \( \pi(a) = \pi(c) = r \quad \text{and} \quad \pi(b) = \pi(d) = s \), for some \( 1 \leq a < b < c < d \leq k \). For a given partition \( \pi \), let \( \{r_1, s_1\}, \ldots, \{r_{\kappa(\pi)}, s_{\kappa(\pi)}\} \) denote the crossings of \( \pi \). As established in Lemma 9 of Ref.~\cite{Erdos2014}, the number of vertices in each intersection \( e_{r_1} \cap e_{s_1}, \ldots, e_{r_{\kappa(\pi)}} \cap e_{s_{\kappa(\pi)}} \) can be treated as approximately independent Poisson random variables with mean \(\lambda_\infty = \lim_{N \to \infty} p^2/N\). Under this assumption, because the normalized trace on a qubit associated with such a twofold crossing \( r \) can be expressed as
\begin{align}
r &= (1 + \varepsilon_x^2 + \varepsilon_y^2)^{-2} \sum_{\alpha, \beta \in \{x, y, z\}} \frac{1}{2} \mathrm{Tr} \tilde{\sigma}^\alpha \tilde{\sigma}^\beta \tilde{\sigma}^\alpha \tilde{\sigma}^\beta \nonumber \\
&= \frac{1 - 2\varepsilon_x^2 - 2\varepsilon_y^2 + \varepsilon_x^4 - 2\varepsilon_x^2 \varepsilon_y^2 + \varepsilon_y^4}{(1 + \varepsilon_x^2 + \varepsilon_y^2)^2},
\end{align}
\( m_k \) is expressed as
\begin{align}
m_k =& \frac{1}{(k/2)!} \sum_{\pi \in \mathcal{S}_k} \sum_{m_1=0}^\infty \cdots \sum_{m_{\kappa(\pi)}=0}^\infty \frac{\lambda^{m_1+\cdots+m_{\kappa(\pi)}}}{m_1! \cdots m_{\kappa(\pi)}!} \nonumber \\
& \times e^{-\kappa(\pi)\lambda_\infty} r^{m_1+\cdots+m_{\kappa(\pi)}} \nonumber \\
=& \frac{(e^{-(1-r)\lambda_\infty})^{\kappa(\pi)}}{(k/2)!}.
\end{align}

Importantly, the generalized expression of $m_k$  retains exactly the same functional form as in the isotropic case. Noting that \( r = -1/3 \) in the isotropic case with \( \varepsilon_x = \varepsilon_y = 1 \), we find that the parameter \( \lambda_\infty \) governing \( \rho_{\lambda_\infty}(E) \) is modified by the introduction of anisotropy and is expressed as
\begin{align}
\tilde{\lambda}_\infty =& \frac{3}{4}(1 - r)\lambda_\infty = \frac{ 3(\varepsilon_x^2 + \varepsilon_y^2 + \varepsilon_x^2 \varepsilon_y^2) }{ (1 + \varepsilon_x^2 + \varepsilon_y^2)^2 } \lambda_\infty.
\label{eq:lambda_tilde_infty}
\end{align}

This parameter \(\tilde{\lambda}_\infty\) represents a rescaled version of \(\lambda_\infty\) that incorporates the contributions from anisotropy in spin space, \( (\varepsilon_x, \varepsilon_y) \neq (1, 1) \). The generalized expression of the density of states remains consistent with the functional form in Eq.~\eqref{eq:rho_lambda_infty}, with the substitution of \(\lambda_\infty\) by \(\tilde{\lambda}_\infty\). Thus, for the $XY$ limit \( (\varepsilon_x, \varepsilon_y) = (1,0) \), taking into account that \(\lambda_\infty/\tilde{\lambda}_\infty = 4/3\) from Eq.~\eqref{eq:lambda_tilde_infty}, the crossover occurs at a $p$ value $2/\sqrt{3}$ times larger than that in the isotropic case as mentioned in Sec.~\ref{sec:XY}.

\bibliography{ref}

\end{document}